\documentclass[useAMS,usenatbib]{mn2e}

\usepackage{epsfig}
\usepackage{aas_macros}
\usepackage{natbib}
\usepackage{times}

\newbox\grsign \setbox\grsign=\hbox{$>$} \newdimen\grdimen \grdimen=\ht\grsign
\newbox\labox \newbox\gabox \newbox\simpropbox \newbox\wtildebox 
\setbox\gabox=\hbox{\raise.5ex\hbox{$>$}\llap
    {\lower.5ex\hbox{$\sim$}}}\ht1=\grdimen\dp1=0pt
\setbox\labox=\hbox{\raise.5ex\hbox{$<$}\llap
    {\lower.5ex\hbox{$\sim$}}}\ht2=\grdimen\dp2=0pt
\def\ga{\mathrel{\copy\gabox}}
\def\la{\mathrel{\copy\labox}}

\newcommand{\be}{\mbox{\begin{equation}}}
\newcommand{\ee}{\mbox{\end{equation}}}

\newcommand{\Cref}{\mbox{$m_{\rm ref}$}}

\newcommand{\msun}{\mbox{M$_\odot$}}

\title{Comparing molecular gas across cosmic time-scales: the Milky Way as both a typical spiral galaxy and a high-redshift galaxy analogue}
\author{J.~M.~Diederik Kruijssen$^1$ and Steven~N.~Longmore$^{2}$\\
$^{1}$Max-Planck Institut f\"{u}r Astrophysik, Karl-Schwarzschild-Stra\ss e 1, 85748 Garching, Germany; kruijssen@mpa-garching.mpg.de\\
$^{2}$Astrophysics Research Institute, Liverpool John Moores University, IC2, Liverpool Science Park, 146 Brownlow Hill,
Liverpool L3 5RF, United Kingdom}

\begin{document}

\date{Accepted 2013 August 5. Received 2013 July 31; in original form 2013 June 26.}

\pagerange{\pageref{firstpage}--\pageref{lastpage}} \pubyear{2012}
\label{firstpage}

\maketitle

\begin{abstract}
Detailed observations of the nearest star-forming regions in the Milky Way (MW) provide the ultimate benchmark for studying star formation. The extent to which the results of these Galaxy-based studies can be extrapolated to extragalactic systems depends on the overlap of the environmental conditions probed. In this paper, we compare the properties of clouds and star-forming regions in the MW with those in nearby galaxies and in the high-redshift Universe. We find that in terms of their baryonic composition, kinematics, and densities, the clouds in the solar neighbourhood are similar to those in nearby galaxies. The clouds and regions in the Central Molecular Zone (CMZ, i.e.~the {inner 250~pc}) of the MW are indistinguishable from high-redshift clouds and galaxies. The {\it presently} low star formation rate in the CMZ therefore implies that either (1) its gas represents the initial conditions for high-redshift starbursts or (2) some yet unidentified process consistently suppresses star formation over $\ga10^8~{\rm yr}$ time-scales. We conclude that the MW contains large reservoirs of gas with properties directly comparable to most of the known range of star formation environments and is therefore an excellent template for studying star formation across cosmological time-scales.
\end{abstract}


\section{Introduction} \label{sec:intro}
Star formation (SF) is {a} fundamental baryonic process in the
Universe, underpinning our understanding of cosmic evolution from the
epoch of re-ionization \citep[e.g.][]{madau99,hopkins06b}, through to the visible structure and chemical
enrichment of galaxies \citep{kennicutt98,evans99}, and down to the formation of planetary
systems \citep{haisch01}. However, despite the conversion of gas into stars being a
cornerstone of astrophysics and cosmology, we lack a clear
understanding of the physics governing this process \citep{kennicutt12}. Developing an
end-to-end model of SF across cosmological time-scales is therefore one
of the key goals in astrophysics today.

Efforts in this direction must be led by observations that can
resolve sites of individual SF {across the full range of protostellar densities, detect the gas mass distribution down to the progenitors of the lowest-mass stars, and also} determine the global properties of
their natal molecular clouds. Given current instrumentational
limitations, the only targets in the Universe for which this is
possible are SF regions in the Milky Way (MW).  For the foreseeable
future, Galactic observational studies must therefore provide the
benchmark for SF theories. However, the scope of these models is not
limited to understanding SF in our own Galaxy. Physical laws are
invariant over space and time. {Barring differences in external influences,} two gas clouds with identical physical properties
should therefore produce the same stellar populations, irrespective of whether
they existed in the early Universe or at the present day. Likewise, it does not matter which physical processes led to the gas having these properties (e.g.~bar inflows, mergers, gravitational (in)stability). Therefore, identifying gas clouds in the Galaxy with properties similar
to those in external systems potentially provides a direct link to
study SF throughout the Universe.

In this paper, we directly compare the properties of different
molecular gas reservoirs in the MW with those in nearby
galaxies and the high-redshift (high-$z$) Universe. Our aim is
to quantify to what extent we can use our own Galaxy as a template for developing an
end-to-end understanding of SF across cosmological time-scales.
We pay particular attention to the Central Molecular Zone of the MW \citep[the CMZ, i.e.~the extreme environment within {250~pc} of
the supermassive black hole at the centre of our Galaxy, see e.g.][]{morris96}, and show that the gas properties there are similar to those in dense,
star-forming galaxies at high $z$. This makes the CMZ an ideal template to study the initial conditions of SF in the early Universe.

\section{Sample}  \label{sec:sample}
\begin{table*}
 \centering
  \begin{minipage}{162mm}
  \caption{Properties of the clouds, regions and galaxies that are considered in this work.}\label{tab:sample}
  \begin{tabular}{@{}l c c c c c@{}}
  \hline 
  Sample group & Sub-sample & Gas tracer & Characteristic velocities$^j$ & Stellar masses & References$^k$\\
  (1) & (2) & (3) & (4) & (5) & (6) \\
 \hline
CMZ & CMZ regions$^a$ & NH$_3(1,1)$ & $\sigma$ (objects), $V_{\rm circ}$ (disc) & y & 1,2,3\\
  & CMZ clouds$^{a,b}$ & HCN (88.632 GHz) & $\sigma$ & n & 4\\\hline
Solar neighbourhood & Perseus clouds and cores$^{b,c}$ & $^{13}$CO$(1,0)$ & $\sigma$ & y & 4\\
 & Galactic clouds$^{c,d}$ & $^{13}$CO$(1,0)$ & $\sigma$ & y & 5\\
 & Galactic cloud centres$^{c,d,e}$ & $^{13}$CO$(1,0)$ & $\sigma$ & y & 5\\\hline
Nearby galaxies & galactic clouds$^f$ & $^{12}$CO$(1,0)$ & $\sigma$ & n & 6\\
 & Antennae clouds$^f$ & $^{12}$CO$(2,1)$ & $\sigma$ & n & 7\\
 & spiral galaxies$^g$ & $^{12}$CO$(2,1)$, $^{12}$CO$(1,0)$ & $V_{\rm circ}^{\rm flat}$ & y & 8\\\hline
High redshift(-like) & Arp~220 regions$^h$ & $^{12}$CO$(1,0)$ & $V_{\rm circ}^{\rm model}$ & y & 9\\
 & high-$z$ clouds & $^{12}$CO$(6,5)$ & $\sigma$ & n & 10\\
 & high-$z$ galaxies$^i$ & $^{12}$CO$(3,2)$, H$\alpha$ & max$(\sqrt{3}\sigma_{\rm int},V_{\rm circ})$ & y & 11,12,13,14\\
\hline
\end{tabular}\\
$^a$Due to extreme opacities and line-of-sight confusion, it is not possible to use available CO data for deriving the necessary quantities.\\
$^b$Gas surface densities are dynamical, because no other gas mass estimates are available.\\
$^c$Stellar masses are estimated using the local stellar mass density in the solar neighbourhood, which is taken to be $\rho_{\rm star}=0.1~\msun~{\rm pc}^{-3}$ \citep[e.g.][]{kuijken89,holmberg00}.\\
$^d$Gas masses may be underestimated by a factor of 2--3 due to the assumption of LTE, and are multiplied by a factor of 2.5 to account for this.\\
$^e$Using the area within the half-intensity isophote of H$_2$.\\
$^f$Only including clouds for which the radius, velocity dispersion, and gas mass are all available. Our dynamical mass estimates are higher by a factor of 1.12 (galactic clouds) or $1.7$ (Antennae clouds) than in the original sources due to using the definition from \citet{bertoldi92}.\\
$^g$Only including the twelve galaxies for which radial mass density profiles of H$_2$ and stars are available. For consistency with the high-$z$ galaxies, H$_2$ and stellar masses are obtained from these profiles by integration to $1.68$ exponential scale lengths of the $\Sigma_{\rm SFR}$ profiles, which corresponds to the half-light radius. The characteristic velocities are taken to be the circular velocities of the flat parts of the rotation curves.\\
$^h$Nuclear components of Arp~220 (west, east, 480-pc disc). The characteristic velocities are obtained from a dynamical model fit to the velocity field \citep{downes98}, and the stellar masses are assumed to be given by the difference between the dynamical and gas masses.\\
$^i$Combined sample of BzK galaxies, sub-millimeter galaxies, etc. In 42 out of 80 galaxies (those from ref.~12), the gas masses had to be derived using H$\alpha$ emission assuming the Schmidt-Kennicutt relation \citep{kennicutt98b} instead of CO. Stellar masses are obtained by spectral fitting.\\
$^j$The characteristic velocity $V$ is taken to be the circular velocity $V_{\rm circ}$ if a region is rotation-dominated (e.g.~galaxy discs), and reflects the velocity dispersion $\sigma$ in dispersion-dominated systems (e.g.~clouds, some of the high-$z$ galaxies).\\
$^k$References: (1) \citet{longmore12}, (2) \citet{longmore13}, (3) \citet{kruijssen13}, (4) \citet{shetty12}, (5) \citet{heyer09}, (6) \citet{bolatto08}, (7) \citet{wei12}, (8) \citet{leroy08}, (9) \citet{downes98}, (10) \citet{swinbank11}, (11) \citet{tacconi08}, (12) \citet{forsterschreiber09}, (13) \citet{tacconi10}, (14) \citet{genzel10}.
\end{minipage}
\end{table*}
Our goal is to compare, as directly as possible, the properties of the gas in SF regions from the MW to the far reaches of the Universe. The ideal data set to address this goal would be homogenous observations probing gas at all distances to similar spatial scales and mass sensitivities, using the same observational tracers of the gas properties. In practice, the extent to which this is possible is limited by the observations of the most distant sources in the sample -- in our case the high-$z$ galaxies. For these sources, measurements of gas masses, velocities, sizes and associated stellar masses exist.

\begin{table}
 \centering
  \begin{minipage}{80mm}
  \caption{Properties of the CMZ regions sub-sample.}\label{tab:cmz}
  \begin{tabular}{@{}l c c c c c@{}}
  \hline 
  Region & $R$ & $V$ & $\Sigma_{\rm gas}$ & $\Sigma_{\rm star}$ & $\Sigma_{\rm dyn}$\\
  (1) & (2) & (3) & (4) & (5) & (6) \\
 \hline
230-pc disc & $230$ & $190$ & $1.2\times10^2$ & $1.2\times10^4$ & $1.1\times10^4$ \\
1.3$^\circ$ cloud & $60$ & $25$ & $2.0\times10^2$ & $1.6\times10^3$ & $3.9\times10^3$ \\
80-pc disc & $80$ & $140$ & $7.5\times10^2$ & $1.9\times10^4$ & $1.8\times10^4$ \\
80-pc ring & $10$ & $38$ & $3.0\times10^3$ & $2.8\times10^3$ & $3.4\times10^3$ \\
The Brick & $2.8$ & $16$ & $5.3\times10^3$ & $6.9\times10^2$ & $1.6\times10^4$ \\
Sgr~B2 & $15$ & $40$ & $8.5\times10^3$ & $2.5\times10^3$ & $5.2\times10^3$ \\
\hline
\end{tabular}\\
$R$ is the radius in units of ${\rm pc}$ (for the 80-pc ring the listed value is the scale height), $V$ is the characteristic velocity (see Table~\ref{tab:sample}) in units of ${\rm km}~{\rm s}^{-1}$, $\Sigma_{\rm gas}$ is the gas surface density, $\Sigma_{\rm star}$ is the stellar mass surface density, and $\Sigma_{\rm dyn}$ is the dynamical mass surface density (see text). Surface densities are listed in units of $\msun~{\rm pc}^{-2}$.
\end{minipage}
\end{table}
In Table~\ref{tab:sample}, we compile a broad range of observations of gas clouds and galaxies that include derived measurements of the above physical properties. The sample choice is aimed to be representative, but not exhaustive, of the known variety of environments. A wide range of cosmic environments necessarily implies the use of a heterogeneous data set, in which different tracers are used to derive the same physical quantities. In the footnotes of Table~\ref{tab:sample} we describe our efforts to remove systematic differences in the assumptions and derivations used to calculate the physical properties in different papers, so we can directly compare the observables between sub-samples.

We make a division into four sample groups, being the CMZ, the solar neighbourhood, nearby galaxies, and high-$z$ galaxies. The latter group includes {both `normal' disc galaxies and starburst systems, as well as} three nuclear regions of the nearby ULIRG Arp~220, which is often considered to be the local analogue of a high-$z$ galaxy \citep[e.g.][]{tacconi08,murray10}. {The merging Antennae galaxies may seem a bit out of place in the nearby galaxy sample, which is dominated by isolated spiral galaxies. However, the figures in \S\ref{sec:plots} show that the Antennae are more reminiscent of nearby spirals than of high-redshift galaxies.} The table also lists the various sub-samples that constitute each group (column~2), the gas tracer that was used to infer the gas properties (column~3), the observable setting the characteristic velocity that is used in this work (column~4), whether or not stellar masses are available (column~5), and the literature references (column~6). Throughout this paper, dynamical masses are defined as $M_{\rm dyn}=5RV^2/G$ for clouds \citep{bertoldi92}, and as $M_{\rm dyn}=RV^2/G$ for galaxies (or their central regions).

We mainly focus on normalized quantities such as gas, stellar and
dynamical surface densities, to avoid any biases introduced by the
wide range of spatial scales covered in our sample. This emphasis on
the properties of the objects themselves (e.g.~their baryonic
composition) allows us to obtain a picture of the initial environment
in which SF proceeds. Other quantities (the star formation rate [SFR],
radiation field, etc.) are discussed in \S\ref{sec:disc}.

The properties of the objects in the CMZ regions sub-sample are summarized in Table~\ref{tab:cmz}, and most of them are obtained from earlier work \citep{longmore12,longmore13,kruijssen13}. The 230-pc and 80-pc discs indicate the central region of the MW averaged over these radii, whereas the properties of the other regions are calculated locally. The stellar surface densities are estimated using the mass profile of \citet{launhardt02}. For the \{1.3$^\circ$~cloud, 80-pc~ring, Brick, Sgr~B2\} this was done at galactocentric radii of $R_{\rm gc}=\{190,80,60,100\}$~pc, while the scale heights from \citet{kruijssen13} were used for estimating the stellar mass in the 230-pc and 80-pc discs. We account for the elongated shape of the Brick (G0.253+0.016) by adopting the virial mass estimate from \citet{longmore12} rather than using the definition from \citet{bertoldi92}.

We note that for the Antennae clouds sub-sample, there is a similar data set based on $^{12}$CO$(3,2)$ measurements available from \citet{ueda12}, which has velocity dispersions of a factor of 2 higher than those obtained by \citet{wei12}. The conclusions below are unaffected by which data set is used.

\section{Scaling relations of clouds and galaxies} \label{sec:plots}
\begin{figure}
\center\resizebox{\hsize}{!}{\includegraphics{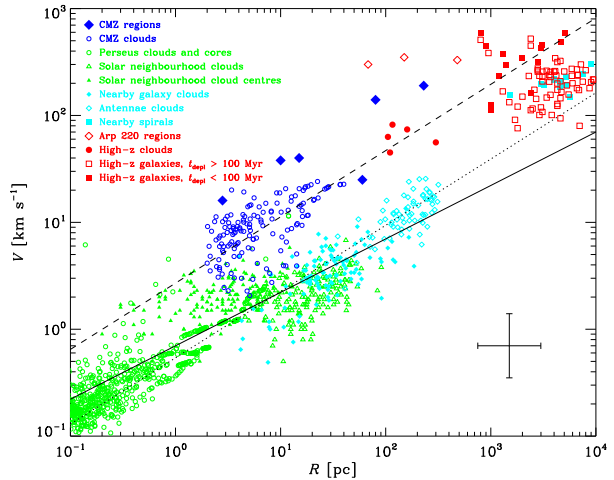}}\\
\caption[]{\label{fig:1}
      {Size-velocity relation of the sample of clouds, regions, and galaxies listed in \S\ref{sec:sample}. The quantity $V$ indicates the characteristic velocity of the system (see \S\ref{sec:sample}). The high-$z$ galaxy sample is divided into two groups of slowly star-forming galaxies (defined as $t_{\rm depl}\equiv M_{\rm gas}/{\rm SFR}>100~{\rm Myr}$) and rapidly star-forming galaxies (defined as $t_{\rm depl}<100~{\rm Myr}$). The error bar indicates a factor-of-two uncertainty, which is characteristic for most of the observations compiled here. The solid line indicates the classical size--linewidth relation of clouds in the solar neighbourhood \citep{solomon87}, the dashed line denotes the power law fit to the CMZ clouds by \citet{shetty12}, and the dotted line shows the same relation shifted down by a factor of 5.}
                 }
\end{figure}
In this section, we systematically compare all of the overlapping properties of the objects in Table~\ref{tab:sample}. Figure~\ref{fig:1} shows the distribution of the data listed in Table~\ref{tab:sample} in the radius-velocity plane, which are the most directly observable quantities available for our sample. Broadly speaking, the objects follow two relations, which likely indicate opposite extremes of a continuum of relations as a result of virial equilibrium (see below). The first roughly matches the classical size-linewidth relation of Galactic clouds \citep{larson81,solomon87}, which is traced by the solar neighbourhood clouds and nearby galaxy clouds (solid line in Figure~\ref{fig:1}). As shown by \citet{heyer09}, a substantial fraction of clouds deviate from the classical relation, because it did not include a dependence on the surface density. A second relation is traced by the CMZ clouds, and is both offset from and slightly steeper than the Larson relation. {The CMZ regions and the} high-$z$ clouds and galaxies with a short gas depletion time-scale follow this `CMZ relation' too. Although it is steeper than the classical relation (with a power law index of 0.7 as opposed to 0.5), shifting the CMZ relation to lower velocity dispersions by a factor of five makes it roughly consistent with the clouds in the solar neighbourhood and nearby spiral galaxies. Interestingly, nearby spiral galaxies and high-$z$ galaxies with a long depletion time-scale lie in between both relations. {However, the classical relation is extrapolated well beyond the fitted range of size-scales -- no kpc-size clouds exist in the local Universe, because the Toomre length is typically only a few 100~pc}. In summary, Figure~\ref{fig:1} suggests that the CMZ clouds and regions as well as the high-$z$ clouds and galaxies are offset to higher $V$ at the same $R$ with respect to clouds in the solar neighbourhood and nearby galaxies. If the objects are in virial equilibrium, then this offset should be correlated with the surface density \citep{heyer09}.

\begin{figure}
\center\resizebox{\hsize}{!}{\includegraphics{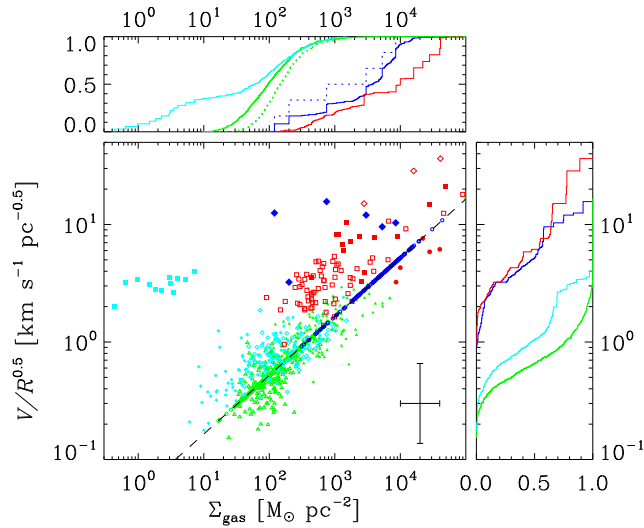}}\\
\caption[]{\label{fig:2}
      {Variation of the scaling coefficient $V/R^{0.5}$ of Figure~\ref{fig:1} with the gas surface density $\Sigma_{\rm gas}$. Symbols and colours are the same as in Figure~\ref{fig:1}, and the dashed line indicates the relation implied by combining two of the \citet{larson81} scaling relations \citep[see text and equation~12 of][]{heyer09}. The CMZ clouds and the Perseus clouds and cores follow the relation by definition, because no gas mass estimates are available. The top and left-hand panels show the cumulative distributions of $\Sigma_{\rm gas}$ and $V/R^{0.5}$, respectively, for each of the four sample groups from Table~\ref{tab:sample}. {The dotted lines in the top panel exclude the CMZ clouds and Perseus clouds and cores, because these sub-samples lack independent gas mass measurements}.}
                 }
\end{figure}
To avoid the clear scale dependence of the quantities shown in Figure~\ref{fig:1}, we now turn to a discussion of normalized observables. Figure~\ref{fig:2} shows the scaling coefficient of the solid line in Figure~\ref{fig:1} as a function of the gas surface density. The cumulative distributions above and to the right of the main panel show how the objects in the different sample groups from Table~\ref{tab:sample} are distributed in the parameter space. To make sure that each sub-sample is clearly visible (i.e.~irrespective of the number of objects it contains), the numbers are rescaled such that each sub-sample contributes an equal fraction to the cumulative distribution of the sample group it belongs to. For instance, the six CMZ regions from Table~\ref{tab:cmz} and the more numerous CMZ clouds both contribute 50\% of the increase of the blue cumulative distribution from zero to unity. The rescaling is visible in the cumulative distributions as discrete steps for small-$N$ sub-samples, and a more gradual increase for large-$N$ sub-samples.

{The gas surface density is the fundamental quantity in many galactic SF relations.} \citet{heyer09} combined the size-linewidth and gravitational equilibrium relations of \citet{larson81} to show that the normalisation of the size-linewidth relation should depend on the surface density $\Sigma$ (or pressure $P\propto\Sigma^2$) as
\begin{equation}
\label{eq:rv}
V=(\pi G/A)^{1/2}\Sigma^{1/2}R^{1/2} ,
\end{equation}
where the $A$ is the numerical factor in the definition of the dynamical mass, i.e.~$A=5$ for clouds and $A=1$ for galaxies (see \S\ref{sec:sample}). This relation is shown in Figure~\ref{fig:2} as the dashed line, assuming that the gas mass dominates the dynamical mass. A large part of our sample indeed follows the relation of equation~(\ref{eq:rv}), and indicates that there exists a continuum of size-linewidth relations in Figure~\ref{fig:1} for different surface densities. The scaling coefficient $V/R^{0.5}$ is proportional to the square-root of the dynamical mass surface density, and hence objects situated above the relation are either supervirial, or have a gravitational potential that is not gas-dominated. The main sub-samples that deviate from the relation are the nearby spiral galaxies, the CMZ regions, and the high-$z$ galaxies. For the spiral galaxies, this is easily understood because their dynamics are dominated by stars and dark matter, while the gas surface density is low because it is averaged over the entire galaxy. For the CMZ regions and high-$z$ galaxies {(including Arp~220)}, the offset is due to the difference in definition of the dynamical mass between clouds and galaxies, as well as due to the presence of stars. Gas-to-stellar mass ratios in the range ${\cal R}_{\rm gas:star}=0.01$--$1$ are needed to move the points onto the relation. Table~\ref{tab:cmz} supports this idea, since only the Brick and Sgr~B2 (the two rightmost blue diamonds in Figure~\ref{fig:2}) are gas-dominated. By contrast, the two top-left blue diamonds represent the CMZ discs, which have higher stellar masses than gas masses. We return to this point below.

\begin{figure}
\center\resizebox{\hsize}{!}{\includegraphics{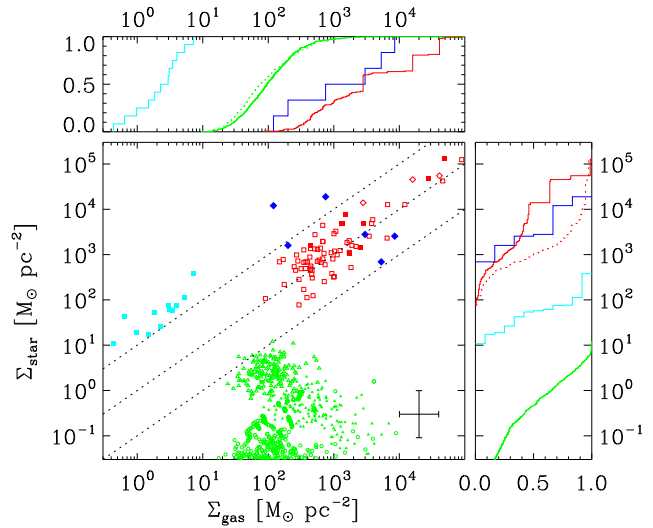}}\\
\caption[]{\label{fig:3}
      {Variation of the stellar surface density $\Sigma_{\rm star}$ with the gas surface density $\Sigma_{\rm gas}$. Symbols and colours are the same as in Figure~\ref{fig:1}, and from top to bottom the dotted lines indicate gas-to-stellar mass ratios of ${\cal R}_{\rm gas:star}\equiv\Sigma_{\rm gas}/\Sigma_{\rm star}=\{0.1,1,10\}$. The top and left-hand panels show the cumulative distributions of $\Sigma_{\rm gas}$ and $\Sigma_{\rm star}$, respectively, for each of the four sample groups from Table~\ref{tab:sample} -- only including objects with stellar mass estimates. {The dotted line in the top panel excludes the Perseus clouds and cores, because for this sub-sample no independent gas mass measurements are available, whereas the dotted red line in the right-hand panel excludes the Arp~220 regions, because the stellar mass estimates are not independent (see the footnote in Table~\ref{tab:sample})}.}
                 }
\end{figure}
The cumulative distributions show that the gas surface densities reached in the CMZ approach those reached in high-$z$ environments {and Arp~220}, and the dynamical surface densities of both sample groups are also similar. This is in stark contrast with the properties of nearby clouds and galaxies. However, no clear separation exists, because both regimes do overlap over a narrow ($<1$~dex) range of surface densities ($\Sigma_{\rm gas}=10^2$--$10^3~\msun~{\rm pc}^{-2}$) and scaling coefficients ($V/R^{0.5}=1$--$4~{\rm km}~{\rm s}^{-1}~{\rm pc}^{-0.5}$).

{Figure~\ref{fig:2} contains several outliers for which most likely the stellar potential is important.} In order to address the gas-to-stellar mass ratios throughout the sample, Figure~\ref{fig:3} shows the stellar surface density as a function of the gas surface density. This necessarily limits the data to those sub-samples for which stellar mass estimates are available (see the fifth column of Table~\ref{sec:sample}). The dotted lines in the main panel of Figure~\ref{fig:3} indicate constant gas-to-stellar mass ratios ${\cal R}_{\rm gas:star}$. Objects near the middle dotted line have comparable gas and stellar masses, while those above (below) are dominated by stars (gas). The solar neighbourhood clouds are naturally gas-dominated, whereas the nearby spiral galaxies have gas fractions of 10\% or lower. By contrast, the CMZ and high-$z$ galaxies have comparable gas and stellar masses with ${\cal R}_{\rm gas:star}=0.1$--$4$. The CMZ points lie around the high-$z$ points with larger scatter -- the two top-left blue diamonds represent the CMZ discs, and the bottom-right diamond indicates the Brick. Figure~\ref{fig:3} confirms the impression from Figure~\ref{fig:2} that the gravitational potentials of the CMZ regions and high-$z$ galaxies have a non-negligible contribution from stars. A similar picture is sketched by the cumulative distributions, which again show that the CMZ regions occupy the same part of parameter space as the high-$z$ galaxies. In terms of the stellar surface densities, there is no real overlap between the local Universe objects and the high-$z$ (analogue) objects. While this may be due to the limitations of the sample, a reasonable division between both regimes would be $\Sigma_{\rm star}\sim2\times10^2~\msun~{\rm pc}^{-2}$.

\begin{figure}
\center\resizebox{\hsize}{!}{\includegraphics{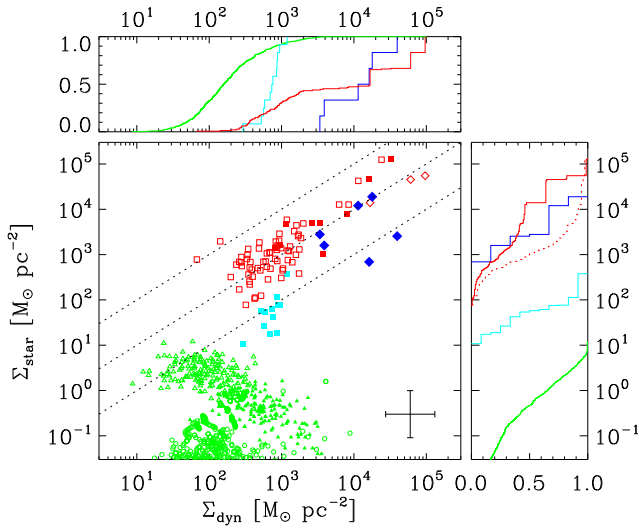}}\\
\caption[]{\label{fig:4}
      {Variation of the stellar surface density $\Sigma_{\rm star}$ with the dynamical surface density $\Sigma_{\rm dyn}$, which assumes virial equilibrium. Symbols and colours are the same as in Figure~\ref{fig:1}, and from top to bottom, the dotted lines indicate stellar-to-dynamical mass ratios of ${\cal R}_{\rm star:dyn}\equiv\Sigma_{\rm star}/\Sigma_{\rm dyn}=\{10,1,0.1\}$. The top and left-hand panels show the cumulative distributions of $\Sigma_{\rm dyn}$ and $\Sigma_{\rm star}$, respectively, for each of the four sample groups from Table~\ref{tab:sample} -- only including objects with stellar mass estimates. {The dotted red line in the right-hand panel excludes the Arp~220 regions, because the stellar mass estimates are not independent (see Table~\ref{tab:sample})}.}
                 }
\end{figure}
Figure~\ref{fig:4} closes the circle of gas, stellar, and dynamical mass densities. It shows the stellar surface density as a function of the dynamical mass surface density, which is proportional to the square of the ordinate of Figure~\ref{fig:2}. Analogously to Figure~\ref{fig:3}, dotted lines indicate constant stellar-to-dynamical mass ratios ${\cal R}_{\rm star:dyn}$. Objects near the middle dotted line have comparable stellar and dynamical masses, while those above are stellar-dominated, and those below either have kinematics dominated by non-stellar mass, or are supervirial. Because their gravitational potential is gas-dominated, the solar neighbourhood clouds and cores have low stellar-to-dynamical mass ratios. The nearby galaxies also have low stellar-to-dynamical mass ratios of ${\cal R}_{\rm star:dyn}\leq0.1$, but for a different reason -- they are dark-matter dominated, because Figure~\ref{fig:3} already indicated that the gas-to-stellar mass ratio is also around ${\cal R}_{\rm gas:star}\leq0.1$. As shown in the discussion of Figure~\ref{fig:2}, the dynamics of the Brick and Sgr~B2 are gas-dominated, and indeed they are represented by the two blue diamonds with high dynamical-to-stellar mass ratios in Figure~\ref{fig:4}. The other CMZ region data points have low gas-to-stellar mass ratios and their alignment with ${\cal R}_{\rm star:dyn}=1$ therefore indicates virial equilibrium. The same holds for the high-$z$ galaxies, which also have comparable stellar and dynamical masses. Combining Figure~\ref{fig:4} with Figure~\ref{fig:2}, we see that the dynamical mass density of all objects in the sample is accounted for by the total baryonic mass (i.e.~they are close to virial equilibrium), except for the nearby spiral galaxies, which are dark-matter dominated. Again, the cumulative distributions show that the CMZ data {occupy} the same part of parameter space as the high-$z$ galaxies. More specifically, there is a low-dynamical surface density tail of high-$z$ galaxies, which is not covered by the CMZ regions sub-sample. The CMZ regions stand clear of the low-$z$ objects, with a separation at around $\Sigma_{\rm dyn}=3\times10^3~\msun~{\rm pc}^{-2}$, and coincide with the extremely high-density, high-$z$ galaxies.

\section{Discussion} \label{sec:disc}
The sample of clouds, regions and galaxies used in this paper is necessarily heterogeneous, and hence there are a number of caveats that require further discussion. Firstly, the tracers of the gas (kinematics) vary between the different sub-samples -- we use NH$_3(1,1)$ and HCN for the CMZ, whereas various CO transitions are used to probe the kinematics in all other samples. This may underestimate the velocity dispersion with respect to CO, because NH$_3$ and HCN trace higher densities (and hence smaller spatial scales). Correcting for this would only exacerbate the difference with clouds in nearby galaxies, which are found to have smaller velocity dispersions than objects in the CMZ. Secondly, the gas masses of some of the high-$z$ galaxies were estimated from the H$\alpha$ flux using the Schmidt-Kennicutt relation. This may overestimate the gas mass in galaxies with short depletion time-scales, which are forming stars at a rate higher than appropriate for the Schmidt-Kennicutt relation \citep{daddi10b}. If true, this would move some of the high-$z$ sample closer to the CMZ regions (see the top panels of Figures~\ref{fig:2} and~\ref{fig:3}). We conclude that the similarity between the CMZ and high-$z$ galaxies is real, irrespective of the heterogeneity.

The CMZ and high-$z$ galaxies also have other properties in common, which we have not compared in detail here. For instance, the high Mach number of the gas in the CMZ \citep[${\cal M}\sim70$, e.g.][]{kruijssen13} is characteristic of starburst and high-$z$ galaxies \citep[${\cal M}\sim100$, e.g.][]{swinbank11}. Due to the similarly high densities ($n\sim10^4~{\rm cm}^{-3}$) and temperatures ($T=50$--$100~{\rm K}$), the turbulent {and thermal} gas pressures in the CMZ and high-$z$ galaxies are comparable, which implies similar surface densities of marginally gravitating clouds in both environments (see Figures~\ref{fig:2} and~\ref{fig:3}). The CO spectral energy distributions should also peak at similar transitions \citep[e.g.][]{danielson11}. Several differences between the CMZ and high-$z$ galaxies exist as well. The metallicity in the CMZ {exceeds that of high-$z$ galaxies} by a factor of a few \citep[e.g.][and references therein]{erb06,longmore13}, whereas the high SFR densities in high-$z$ galaxies imply vastly greater radiation and cosmic ray pressures than in the CMZ.

The most notable difference between the CMZ and the vigorously star-forming (i.e.~sub-mm) galaxies is that the former is forming stars at a rate 1--2 orders of magnitude below the predictions of galactic scaling relations \citep{longmore13}. This is unlikely to be caused by the aforementioned differences. A higher metallicity should imply more efficient cooling and more efficient SF, contrary to the low SFR that is observed in the CMZ. The elevated radiation and cosmic ray pressures in high-$z$ galaxies are caused by SF itself, and it would therefore be a circular argument to attribute a dearth of SF to a low feedback pressure. A more likely explanation is that the CMZ is currently in a pre-starburst phase and is expected to form stars at a comparable SFR density to high-$z$ galaxies within the next (few) 10~Myr \citep{kruijssen13}. In this scenario, the CMZ provides the initial conditions of high-$z$ starbursts. Alternatively, some yet unidentified process consistently suppresses SF in the CMZ, opening up a new window to understanding galactic SF.

In conclusion, we find that in terms of their baryonic composition, kinematics, and densities, the clouds and regions of the CMZ appear indistinguishable from high-$z$ clouds, as well as high-$z$ galaxies as a whole when scaled to a similar size. Hence, the CMZ provides insight into the initial conditions of SF at high $z$. We show that the properties of the high-density clouds in the CMZ are very similar to the high-$z$ clouds \citep[also see][]{swinbank11}, even more so than the regions of the classical high-$z$ analogue Arp~220. The CMZ clouds may therefore be used as templates to make predictions for the evolution (and possibly the life cycle) of high-$z$ clouds. We propose that the CMZ replaces Arp~220 as the nearest high-$z$ galaxy analogue. In combination with the more quiescent environments of the solar neighbourhood and the Galactic disc, this implies that a remarkably broad range of environments can be probed within our Galaxy alone, making it an exquisite template for obtaining an end-to-end understanding of the SF process.

\section*{Acknowledgments}
We thank Rahul Shetty for providing the data from \citet{shetty12} in electronic form. We are grateful to an anonymous referee, Luca Cortese, Bruce Elmegreen and Mark Swinbank for helpful comments. We acknowledge the hospitality of the Aspen Center for Physics, which is supported by the National Science Foundation Grant No.~PHY-1066293.

\bibliographystyle{mn2e}
\bibliography{mybib}

\bsp

\label{lastpage}

\end{document}